\documentclass[prb,twocolumn,showpacs]{revtex4}

	\usepackage{graphicx,amsmath,amssymb,txfonts}

\newcommand{\bq}{{\bf q}}
\newcommand{\bk}{{\bf k}}
\newcommand{\br}{{\bf r}}

\newcommand{\Hmath}{\mathcal{H}}
\newcommand{\Mmath}{\mathcal{M}}
\newcommand{\ua}{\uparrow}
\newcommand{\da}{\downarrow}
\newcommand{\beq}{\begin{equation}}
\newcommand{\beqn}{\begin{eqnarray}}
\newcommand{\eeq}{\end{equation}}
\newcommand{\eeqn}{\end{eqnarray}}
\newcommand{\nn}{\nonumber}

\begin{document}

	\title{The plasma picture of the fractional quantum Hall effect with internal SU($K$) symmetries}
	
	\author{R. de Gail$^1$, N. Regnault$^2$, and M. O. Goerbig$^1$, }

	\affiliation{
$^1$Laboratoire de Physique des Solides, CNRS UMR 8502, Univ. Paris-Sud, F-91405 Orsay cedex, France\\
$^2$Laboratoire Pierre Aigrain, D\'epartement de Physique, Ecole Normale Sup\'erieure, 24 Rue Lhomond, F-75005 Paris, France}


	\begin{abstract}
We consider trial wavefunctions exhibiting SU($K$) symmetry which may be well-suited to grasp the physics of the fractional quantum Hall effect with internal degrees of freedom. Systems of relevance may be either spin-unpolarized states ($K=2$), semiconductors bilayers ($K=2,4$) or graphene ($K=4$).
We find that some introduced states are unstable, undergoing phase separation or phase transition. This allows us to strongly reduce the set of candidate wavefunctions eligible for a particular filling factor.
The stability criteria are obtained with the help of Laughlin's plasma 
analogy, which we systematically generalize to the multicomponent SU($K$) case.
The validity of these criteria are corroborated by exact-diagonalization studies,
for SU($2$) and SU($4$).
Furthermore, we study the pair-correlation functions of the ground state
and elementary charged excitations within the multicomponent plasma picture.

	\end{abstract}


	\maketitle


	\section{Introduction}


Soon after the discovery of the fractional quantum Hall effect
(FQHE), \cite{Tsui} Laughlin successfully described the underlying 
strongly-correlated electron liquid with the help of a simple trial 
wavefunction. \cite{Laughlin} The reasons for the success of this approach 
were twofold: first, the calculated energy of this state is lower than
that of charge-density waves or Wigner crystals,\cite{fukuyama}
 which are natural candidates
for the ground state within the partially filled lowest Landau level (LL) 
due to the quenched kinetic energy.\cite{Laughlin} The second reason
for its success is the fact that Laughlin's wavefunction is 
in very sharp agreement with exact-diagonalization studies.\cite{EDlaughlin}

A powerful tool in the understanding of Laughlin's wavefunction is 
the quantum-classical analogy, in which its probability is interpreted as 
the (classical) Boltzmann weight of a two-dimensional (2D) one-component plasma
(2DOCP).\cite{Laughlin} Most strikingly, this analogy shows that Laughlin's
trial wavefunctions have no free parameter to be optimized by any variational 
calculation.

Laughlin's original proposal was concerned with only one single species of
fermions, namely spin-polarized electrons. In spite of its success, this 
is at first sight a very crude assumption in view of the relatively weak
(effective) Zeeman effect when compared to the leading energy scale set
by the Coulomb interaction $e^2/\epsilon l_B$, in terms of the magnetic
length $l_B=\sqrt{\hbar/eB}$. Indeed the latter is almost two orders of
magnitude larger than the bare spin splitting for typical magnetic fields of
$B\sim10$ T. In order to account for an
internal SU(2) spin symmetry, Halperin proposed a generalized 
trial wavefunction,\cite{Halperin} which includes Laughlin's as a special case.
The latter may indeed be viewed as a Halperin wavefunction with a spontaneous
ferromagnetic spin ordering.\cite{DasSarma} 

Halperin's SU(2) wavefunctions have been a first step in the understanding of
general multicomponent systems. In the case of bilayer quantum Hall systems,
the same wavefunctions may be applied if one supposes a complete polarization
of the physical spin and if one interprets the two layer indices as the two
possible orientations of a {\sl pseudospin}.\cite{DasSarma,Moon}
However, the hypothesis of complete spin polarization is {\sl a priori} as 
feably justified in bilayer as in monolayer quantum Hall systems, 
again due to a relatively weak Zeeman effect. A more appropriate approach
is therefore one that takes into account the internal SU(4) spin-pseudospin 
symmetry. Such approaches have indeed been proposed in the description
of ferromagnetic states when the LL filling factor $\nu=n_{el}/n_B$, in terms
of the electronic, $n_{el}$, and the flux, $n_B=eB/h$, densities, respectively,
is $1,2$, or $3$.\cite{ezawa,arovas} 

Another example of a multicomponent quantum Hall system is graphene, where 
the internal SU(4) symmetry is due to the physical spin accompanied by a
twofold valley degeneracy.\cite{GrapheneExp,GrapheneTheo,yang2} In contrast to
the abovementioned bilayer quantum Hall systems, where the pseudospin
symmetry is explicitely broken because of the difference between intra- and
interlayer Coulomb interactions, the SU(4) symmetry is almost perfectly
preserved in graphene from an interaction point of view -- a possible 
(valley) symmetry breaking may be due to lattice effects, which are suppressed
by the small parameter $a/l_B$, where $a=0.14$ nm is the distance
between nearest-neighbor carbon atoms in graphene, as compared to
$l_B=26\sqrt{B{\rm [T]}}$ nm.\cite{Goerbig_Graphene,AF,Fuchs,abanin,Herbut} 
In order to describe a possible, yet unobserved, FQHE in graphene, taking
into account the appropriate form of the interaction potential,
\cite{Nomura,Goerbig_Graphene} exact diagonalization studies have been
performed in the framework of an internal SU(2) valley 
symmetry,\cite{Apalkov,Toke} as well as in a SU(4) composite-fermion 
approach.\cite{yang,toke} 

More recently, two of us have proposed a generalization of Halperin's 
wavefunctions to $K$ components, i.e. systems with an internal SU($K$)
symmetry, in order to describe a possible FQHE in $K$-component systems,
namely graphene with $K=4$.\cite{Goerbig_SU4} Here, we investigate 
the stability of these wavefunctions from two complementary perspectives 
-- first, we 
derive stability criteria within a generalized plasma picture. This
analogy allows one to interpret the SU($K$) Halperin wavefunctions in terms
of $K$ correlated 2DOCP and to describe in a compact manner their ground-state
properties as well as the elementary excitations with fractional charge.
In a second step, we corroborate the validity of the generalized plasma 
picture with the help of exact-diagonalization studies.

After a brief review of Laughlin's plasma analogy (Sec. II), we generalize
the plasma picture to $K$-component systems in Sec. III. In Sec. IV, we
derive general stability criteria within the plasma analogy, on the basis
of which we discuss the stability of specific SU(2) and SU(4) wavefunctions.
We complete this paper with a discussion of ground-state properties, such
as sum rules for the pair correlation functions (Sec. V), and fractional charges of
quasiparticle/-hole excitations (Sec. VI).

	\section{Laughlin's Plasma Analogy}

In order to describe a correlated electron liquid to account for the FQHE, 
Laughlin proposed the $N$-particle trial wavefunction\cite{Laughlin}
\beq\label{eq01}
\Psi_m\left(\left\{z_k\right\}\right)=\prod_{k<l}^N\left(z_k-z_l\right)^m 
\exp\left(-\sum_k^N\frac{|z_k|^2}{4}\right),
\eeq
where $z_k=x_k+iy_k$ denotes the position of the $k$-th electron in the complex
plane. Here and in the following, we set the magnetic length $l_B\equiv 1$, for 
notational convenience. The form of this trial wavefunction is solely determined
by the analyticity condition for the lowest LL -- i.e. all single-particle states 
are of the form $z^{\ell}\exp(-z^2/4)$, where $\ell$ is a positive integer 
-- and by symmetry considerations. In order to 
have a translational and rotational invariant state
and thus an incompressible state with no gapless
Goldstone mode, the wavefunction may only 
depend on the relative distance $z_k-z_l$ of the $k$-th and the $l$-th particle. 
Furthermore, fermion statistics for electrons requires the exponent $m$ to be an 
odd integer, which is the only variational parameter of Laughlin's wavefunction
(\ref{eq01}).

However, the parameter $m$ turns out to be fixed by the electon density, or else the filling factor, $\nu=1/m$, as Laughlin showed with the help of a plasma 
analogy.\cite{Laughlin}
Indeed, one may interpret the modulus square of the wave function as the 
Boltzmann weight 
\beq\label{eq02}
\left|\Psi_m\left(\left\{z_k\right\}\right)\right|^2=e^{-\beta\Hmath_N}
\eeq
of a classical system, namely a 2DOCP described by
the classical Hamiltonian\cite{Laughlin,OCP,Bhatta}
\beq\label{eq03}
\Hmath_N=-m\sum_{k<l}\ln|z_k-z_l|+\sum_k\frac{|z_k|^2}{4},
\eeq
where one has set somewhat arbitrarily the inverse "temperature" $\beta\equiv 2$.
Notice that the true temperature does not intervene in the analysis because the
system is placed at $T=0$. 
The first term describes 2D interacting particles of charge $\sqrt{m}$, whereas
the second term may be interpreted as a homogeneous background of charge
$-1/\sqrt{m}$ (jellium).
The minimization of the classical Hamiltonian 
corresponds, via the relation (\ref{eq02}), to a maximal quantum probability of the
original quantum
system of electrons within the lowest LL. The classical ground state of
the Hamiltonian (\ref{eq03}), however, is obtained when the plasma particles of 
charge $\sqrt{m}$ are fully neutralized by the background, i.e. when
\beq\label{eq04}
m n_{el}=\frac{1}{2\pi} \qquad \Leftrightarrow \qquad m\nu=1.
\eeq
It is evident from the last equation that the variational parameter must be 
positive -- otherwise one would have to treat with unphysical negative densities.
Notice that from the wavefunction point of view, $m<0$ is not physical because it
violates the analyticity condition for
wavefunctions in the lowest LL. This point, which may seem obvious, is worth  
being emphasized and being recalled in the following sections when the plasma 
picture is generalized to more components.

		\section{Plasma Picture for SU($K$) Halperin wavefunctions}

Based on Halperin's idea to write down a Laughlin-type wavefunction for 
a two-component quantum Hall system, in order to take into account the spin degree
of freedom,\cite{Halperin} two of us have proposed a SU($K$) generalization for
a $K$-component system,\cite{Goerbig_SU4}
\beqn\label{wave}
\nn
\Psi^{SU(K)}_{m_1,...,m_K;n_{ij}} &=&  \Phi^L_{m_1,...,m_K}\times \Phi^{inter}_{n_{ij}} \times \exp{ \left( \displaystyle{-\sum_{i=1}^{K} \sum_{k_i = 1}^{N_i} \frac{|z_{k_i}^{(i)}|^2}{4}} \right)} \\
\Phi^L_{m_1,...,m_K}&=&\displaystyle{\prod_{i = 1}^{K}  \prod_{k_i < l_i}^{N_i} (z_{k_i}^{(i)} - z_{l_i}^{(i)})^{m_i}} \\
\nn
\Phi^{inter}_{n_{ij}}&=&\displaystyle{\prod_{i < j}^{K} \prod_{k_i = 1}^{N_i} \prod_{k_j = 1}^{N_j} (z_{k_i}^{(i)} - z_{k_j}^{(j)})^{n_{ij}}} \,.
\eeqn
There are $K$ different types of electrons [denoted with superscript $(i)$] with inter- ($n_{ij}$) and intra-component ($m_{i}$) quantum correlations, and
 $z_{k_i}^{(i)}$ is the complex position of $k_i$-th electron of type
$i=1,...,K$. The lowest-LL analyticity condition imposes that all exponents,
$m_i$ and $n_{ij}$, must be integers. Furthermore, $m_i$ must be odd in the case of
fermions. Apart from $K=2$, discussed by Halperin,\cite{Halperin} 
$K=4$ wavefunctions may be physically significant in the case of bilayer quantum
Hall systems and graphene. In the former example, the internal degrees of freedom
do not only contain the physical SU(2) spin ($\ua,\da$), but also a layer index, 
which may be mimicked by an additional SU(2) isospin ($+,-$). There are thus four
internal states, $1=(\ua,+)$, $2=(\ua,-)$, $3=(\da,+)$, and $4=(\da,-)$. In the
case of graphene, an isospin ($+,-$) must be introduced in order to account for
the two-fold valley degeneracy. Wavefunctions similar to those in Eq. (\ref{wave})
have been proposed by Qiu {\sl et al.} for multilayer quantum Hall 
systems,\cite{Qiu}
by Morf as potential candidates for the FQHE hierarchy states,\cite{morf}
and by Yang {\sl et al.} in the study of a possible FQHE in graphene.\cite{yang}

Again, the starting point of the plasma analogy is Eq. (\ref{eq02}), and one 
associates the new Hamiltonian $\Hmath_N$ with a physical system.
In the case $K=2$, this system has been interpreted as a generalized plasma, which consists of $K$ different particle types (each of which corresponds to a different electron type in the original quantum system) plus a neutralizing background.\cite{Girvin_Correlation,DasSarma} 
With the identification (\ref{eq02}), one obtains for the wavefunctions (\ref{wave})
the classical Hamiltonian
\beqn\label{eq05}
\nn
\Hmath_N &= & -  \sum_{i = 1}^K m_i \sum_{k_i < l_i}^{N_i}  \ln|z_{k_i}^{(i)} - z_{l_i}^{(i)}|  - \sum_{i < j}^K n_{ij} \sum_{k_i, \, k_j}  \ln|z_{k_i}^{(i)} - z_{l_j}^{(j)}| \\
&&+ \sum_{j = i}^K \sum_{k_i = 1}^{N_i} \frac{|z_{k_j}^{(i)}|^2}{4}\ .
\eeqn
Here the first term represents a sum over $K$ 2D interaction terms for $(i)$-type particles of charge $\sqrt{m_i}$, whereas the second one takes into account 
interactions between particles of different type, $(i)$ and $(j)$. However, this 
generalized plasma does not satisfy the charge superposition principle\cite{Jackson} unless
$n_{ij}=\sqrt{m_im_j}$, which is a rather special case. 

Instead of one single plasma of $K$ types of particles, it seems therefore more
appropriate to
interpret this generalized plasma in terms of $K$ different 2DOCPs (one for each type of electrons) with  correlations between them. For this purpose, we introduce the continuum limit, in which the density for particles of type $(i)$ (electrons or plasmatic particles) is $\rho_i(\mathbf{r})=\sum_{k_i} \delta(\mathbf{r-r_{k_i}})$. In order to distinguish the resulting Hamiltonian from the original discrete one, we 
supress the $N$ subscript in Eq. (\ref{eq03}), and one obtains the energy functional
\beq\label{eq06}
\begin{array}{ll}
\Hmath[\{ \rho_i(\mathbf{r}) \}] = &
\displaystyle{ -\iint_{\Omega} d^2r \, d^2r'
 \begin{pmatrix}
   \rho_1 (\mathbf{r}) \\
   \vdots  \\
   \rho_K (\mathbf{r})
 \end{pmatrix}^{\top}
\frac{M_K}{2}
\ln |\mathbf{r} - \mathbf{r'}|
 \begin{pmatrix}
   \rho_1 (\mathbf{r'}) \\
   \vdots  \\
   \rho_K (\mathbf{r'})
 \end{pmatrix}}
\\
& +\displaystyle{\int_{\Omega} d^2r
 \begin{pmatrix}
   \rho_1 (\mathbf{r}) \\
   \vdots  \\
   \rho_K (\mathbf{r})
 \end{pmatrix}^{\top}
\frac{|r|^2}{4} 
\begin{pmatrix}
   1\\
   \vdots  \\
   1
 \end{pmatrix}\ .
}
\end{array}
\eeq
Here, $M_K$ is the symmetric exponent matrix, with $n_{ij}=n_{ji}$ and 
$n_{ii}\equiv m_i$,\cite{Goerbig_SU4} and $\Omega$ is the surface occupied by
the plasma. Similarly to the one-component case, the configurations with maximal 
probability [Eq. (\ref{eq02})] are obtained by
minimizing $\Hmath$ with respect to all densities. The
stationary points are found at
\beq\label{min}
\left.\frac{\delta \Hmath}{\delta \rho_i(\mathbf{r})}\right|_{\rho_{j, \, j\neq i}} = 0.
\eeq
In order to have a minimum, the Hessian matrix
\beq\label{Hesse}
\frac{\delta^2 \Hmath}{\delta \rho_i(\mathbf{r})\delta \rho_j(\mathbf{r})}  \propto M_K,
\eeq
which is identical to the exponent matrix $M_K$ up to a positive constant, needs
to be positive, i.e. have positive eigenvalues.
One may interpret Eq. (\ref{min}) as the stationary point of a 2DOCP of ($i$)-type
particles, whereas the positions of all other particles of type ($j \neq i$) are 
fixed and constitute a quasi-static impurity potential felt by the ($i$)-type 
particles. For the 2DOCP of this type, the interactions between all other types
of particles yields only an unimportant constant, with respect to the 
$\rho_i$ derivative. In this sense, one may indeed interpret the system as $K$ 
correlated 2DOCPs rather than a single plasma of $K$ different types of
particles.

In the same manner as for a single 2DOCP, Eq. (\ref{min}) is satisfied when each of the $K$ plasmas exhibits quasi-neutrality,\cite{Bhatta} but now contributions from the impurities have to be taken into account,
\begin{subequations}
\beqn\label{ff}
m_i\rho_i(\br)+\sum_{j\neq i}n_{ij}\rho_j(\br) = \frac{1}{2\pi}\\
\label{ffm}
\Leftrightarrow
M_K\left(\begin{array}{c}
\rho_i(\br) \\ \vdots \\ \rho_K(\br)
\end{array}\right)=\frac{1}{2\pi}\left(\begin{array}{c}
1 \\ \vdots \\ 1
\end{array}\right).
\eeqn
\end{subequations}
Here, $1/2\pi$ on the r.h.s. of Eq. (\ref{ff}) represents the neutralizing background, as for a single 2DOCP [Eq. (\ref{eq04})].
The second term in Eq. (\ref{ff}) represents the contributions from type-($j \neq i$)
particles due to inter-component correlations. 
One notices that Eq. (\ref{ffm}) is the matrix generalization of Eq. (\ref{eq04}).
This result was previously derived by counting the zeros of the wavefunction
(\ref{wave}).\cite{Goerbig_SU4}  Invertible matrices yield a unique solution with all densities being uniform, $\rho_i(\br) = \rho_i$, as it is the case for U($1$) Laughlin's liquid. The case of non-invertible matrices will be discussed in the next section.

Unlike the U($1$) case, fixing the total filling factor $\nu_T=\nu_1+...+\nu_K$
does not determine uniquely the exponent matrix. 
It has been pointed out in Ref. \onlinecite{Goerbig_SU4} that several candidate
wavefunctions may give rise to a FQHE at the same filling factor $\nu_T$, especially
in the case of larger internal symmetry groups. Moreover, even if one fixes all
component filling factors $\nu_i=2\pi\rho_i$, 
the wavefunction is not unambiguous. As an
example, we consider the SU($2$) Halperin wavefunctions ($m_1,m_2,n$), with (i)
$m_1=m_2=3$, $n=1$ [(331) wavefunction], and (ii) $m_1=m_2=1$, $n=3$ 
[(113) wavefunction]. Both wavefunctions describe a situation with 
$\nu_1=\nu_2=1/4$ and have been considered in the past within the study of a 
possible unpolarized $\nu_T=1/2$ state.\cite{Yoshioka} Indeed, an even-denominator
quantum Hall state has been observed at $\nu=5/2$ and $\nu=7/2$, in the first
excited LL.\cite{5_2} However, it is strongly unlikely that this state is spin-unpolarized\cite{morf2,pan,halperin07}
and more sophisticated theories, in terms of a Pfaffian state, need to be invoked
to account for a spin-polarized state.\cite{theory5_2} 
As we will show below, stability conditions related to Eq. (\ref{Hesse}) allow 
one, in the case of unpolarized states, to discriminate between ($mmn$) and
($nnm$) wavefunctions, for $m>n$.

		\section{Stability}
In order to obtain a stable state, the stationary point obtained from Eq. 
(\ref{min}) must be a minimum, i.e. the Hessian matrix $\propto M_K$ in
Eq. (\ref{Hesse})
must be positive. Otherwise, the plasmas would not be stable, or else in the state of lowest energy. Hence, a first stability condition imposes that all
eigenvalues $\lambda_i$ of $M_K$ must be positive. 

There is indeed a limiting case for which one, or more, eigenvalue(s) is (are) zero. Because the potential is quadratic, the minimum point becomes now a line of minima, which correspond to different ground states. If at least one eigenvalue
is zero, the exponent matrix $M_K$ is no longer invertible, and
all densities may not be fully determined from Eq. (\ref{ffm}).

This point may alternatively be interpreted in terms of SU($K$) ferromagnets -- 
different states of equal energy may, e.g., occur at various combinations of
two (or more) filling factors $\nu_i$ and $\nu_j$, although the sum $\nu_{ij}=
\nu_i+\nu_j$ is fixed. In this case one may introduce a pseudo-spin operator
$S_{ij}^z=N(\nu_i-\nu_j)/2$, which can possibly take all values in between 
$-N/2\leq S_{ij}^z\leq N/2$. The simplest example of such a case is the Laughlin
wavefunction with an internal spin degree of freedom [a Halperin ($mmm$)
wavefunction], which is, for odd $m$, completely antisymmetric in its orbital part.
For fermions, the spin wavefunction must therefore be completely symmetric and represents thus a SU($2$) ferromagnet. If the total spin is 
oriented along the positive $z$ direction, all electrons reside in the upper spin
branch ($\nu_1=1/m$, $\nu_2=0$). In the absence of a Zeeman effect, 
this state has the same energy as the one with a total spin in the $-z$ 
direction ($\nu_1=0$, $\nu_2=1/m$), as well as any intermediate state with $\nu_1+\nu_2=1/m$. 
In the general SU($K$) case, the ferromagnetic properties are determined by the
rank $r$ of the matrix $M_K$. Indeed, if $r<K$ and one introduces common fixed
filling factors $\nu_{ij}$ for the relevant components, one may describe the 
resulting state by a SU($r$) Halperin wavefunction with an invertible exponent
matrix $\tilde{M}_r$ with additional pseudo-spin degrees of freedom for the 
components the density of which remain undetermined.\cite{Goerbig_SU4}

As mentioned earlier, some exponent matrices can lead to negative density solutions for Eq. (\ref{ffm}). A second class of stability conditions needs to be imposed in order to prevent this unphysical situation,
which may occur even in the case of a positive matrix $M_K$.

In order to illustrate the two conditions, we discuss some specific examples for different $K$. The case $K=1$ has already been presented above.

		\subsection{The case $K=2$}
 
We first study Halperin's wavefunction ($m_1, m_2, n$) for the SU($2$) case, which is described by the exponent matrix 
$$M_{K=2} = 
\begin{pmatrix}
m_1 & n \\
n & m_2 \\
\end{pmatrix}.
$$
Even if all exponents are positive, as required by the lowest-LL analyticity
condition, the eigenvalues $\lambda_{\pm}$ and the filling factors $\nu_{1/2}$ are not necessarily so,
\begin{subequations}
\beqn\label{eigen2}
\displaystyle{ \lambda_{\pm} = \frac{m_1 + m_2 \pm \sqrt{(m_1 - m_2)^2 + 4n^2} }{2} }\\
\label{ff2}
\displaystyle{ \nu_1 = \frac{m_2 - n}{m_1 m_2 - n^2} \quad  \nu_2 = \frac{m_1 - n}{m_1 m_2 - n^2} }.
\eeqn
\end{subequations}
In order to obtain only positive eigenvalues (first stability condition), one 
needs to require
\beq\label{eigenc2}
m_1 m_2 - n^2 \geq 0.
\eeq
The case $m_1m_2=n^2$ corresponds to a situation of a non-invertible matrix of 
rank $r=1$. Because of Eq. (\ref{eigenc2}), positive
densities (filling factors) are found from Eq. (\ref{ff2}) only for 
\beq\label{ffc2}
m_1 \geq n \quad {\rm and} \qquad m_2 \geq n,
\eeq
which we thus need to impose as a second class of stability conditions. One furthermore notices form Eqs. (\ref{eigenc2}) and (\ref{ffc2}) that the only states with a non-invertible exponent matrix of rank $r=1$ are the 
ferromagnetic Laughlin states ($mmm$) discussed above.

\begin{figure}
\centering
\includegraphics[width=8cm,height=6cm]{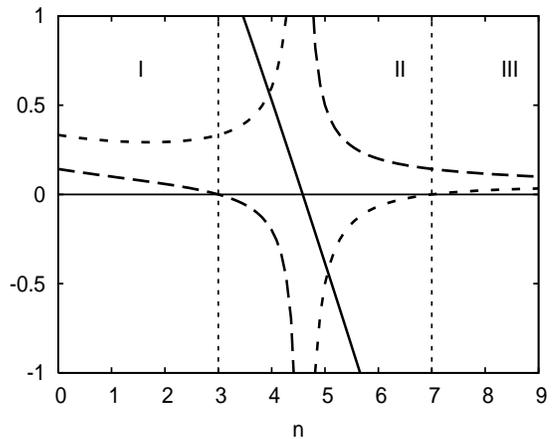}
\caption{\footnotesize Stability of the ($73n$) wavefunction. Both filling factors, $\nu_1$ (long dashed line) and $\nu_2$ (short dashed line), and the $\lambda_{-}$ eigenvalue (solid line) of the ($73n$) wavefunction are plotted as a function of $n$. For $n \leq 3$ (part I) all quantities are positive and the corresponding state is stable. In part II, one of the filling factors is negative. Part III exhibits positive filling factors but the state is still unstable because of the negative eigenvalue; the system eventually undergoes a phase separation.}
\label{73n}
\end{figure}

\begin{figure}
\centering
\begin{tabular}{p{4cm} p{4cm}}
{(a)}
\includegraphics[width=4cm,height=3cm]{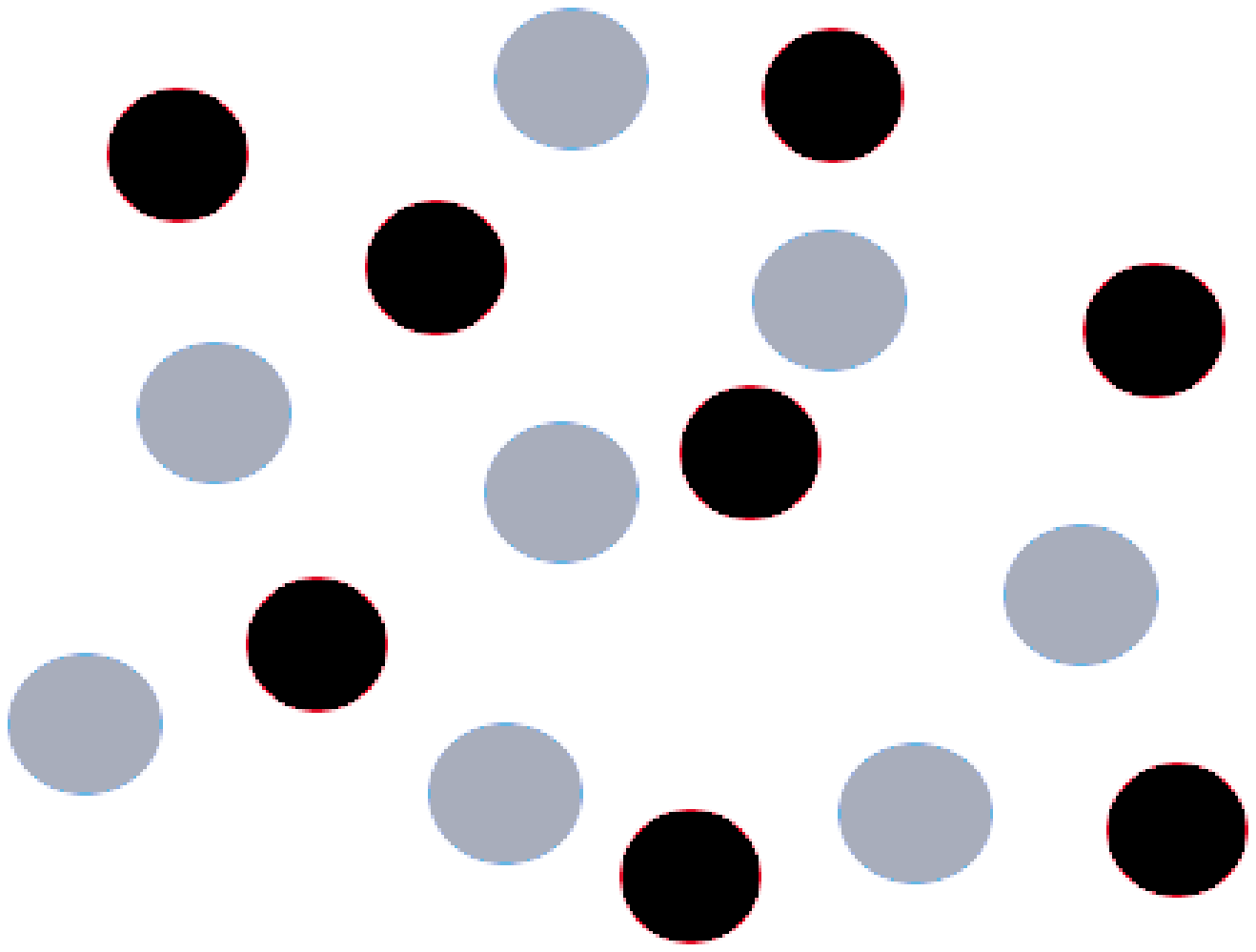}
&
{(b)}
\includegraphics[width=4cm,height=3cm]{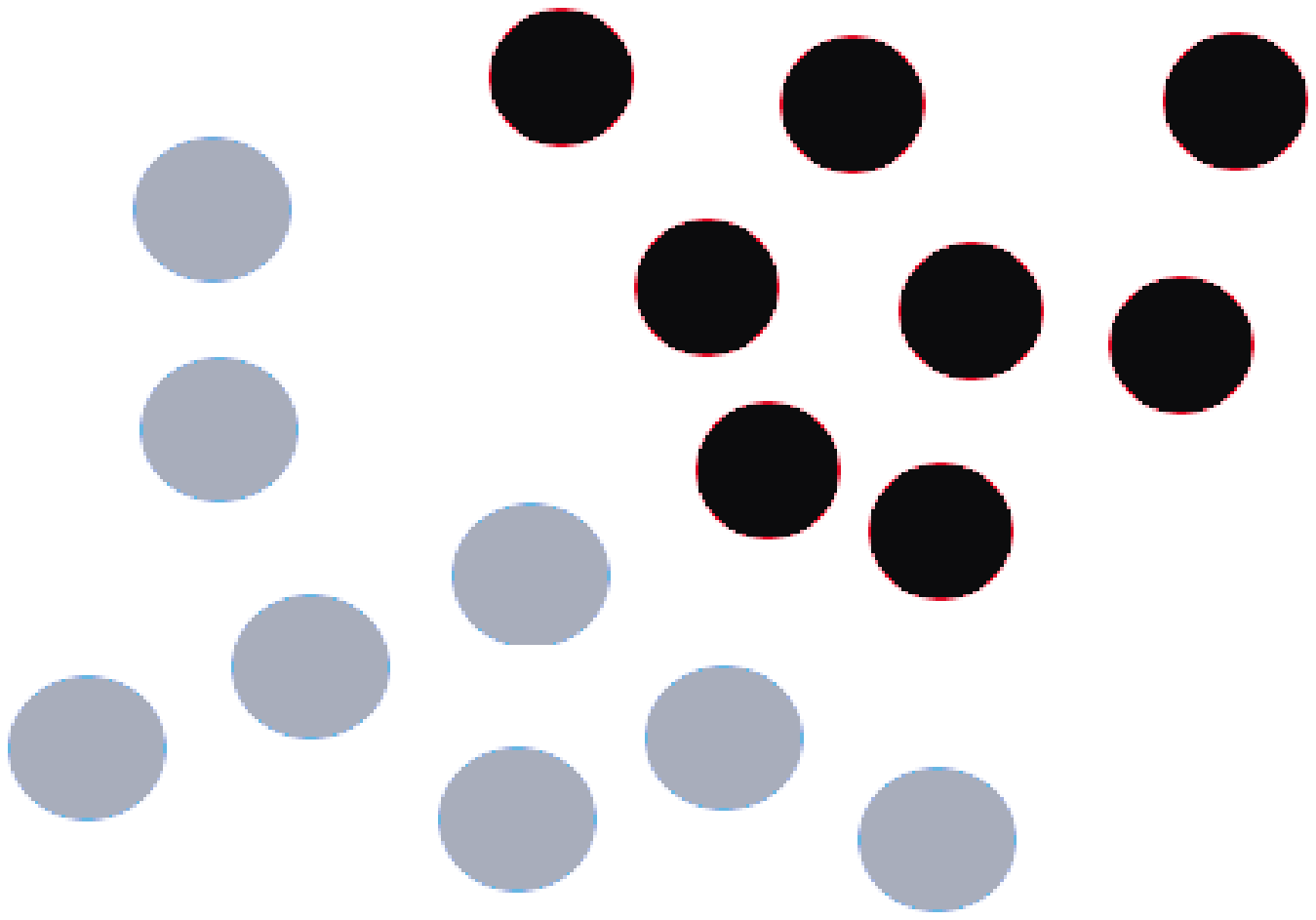}\\
\end{tabular}
\caption{\footnotesize Sktech of the plasma corresponding to the
($73n$) wavefunction, for {\sl (a)} $n\leq 3$ and {\sl (b)}
$7\leq n$. {\sl (a)} Intra-component are stronger than inter-component repulsions ($n\leq m_1,m_2$). Type-($1$) particles (black) are therefore on the average surrounded by type-(2) particles (grey). This yields a stable state of
two homogeneous interpenetrating plasmas. {\sl (b)} If the inter-component
is stronger than the intra-component repulsion, the plasmas have a tendency to 
phase-separate to minimize the number of neighbors from different types. 
}
\label{n<3}
\end{figure}

The final stability criterion Eq. (\ref{ffc2}) for SU($2$) wavefunctions has a
compelling physical interpretation: {\sl intra-component must always be stronger 
than inter-component correlations}. Within the plasma picture, this may also be understood from Figs. \ref{73n}, \ref{n<3}(a), and \ref{n<3}(b).
For illustration, we consider the ($73n$) wavefunction, where $n$ is left as a variable, which we treat in a rather artificial manner as a continuous variable
in the following discussion. It is evident that only integer values may be 
taken into account for physical candidate wavefunctions.

Fig. \ref{73n}  shows the plot of both component filling factors and the lower eigenvalue, $\lambda_{-}$. This graph can be split into three distinct parts I,II, and III.

\begin{itemize}
\item {\sl Part I ($n \leq 3$).} Each type-($1$) particle carries a charge 
$\sqrt{7}$ and is affected by those from the type-($2$) plasma through a 
charge coupling of $n/\sqrt{7}\leq \sqrt{7}$, which may be interpreted 
as constituting a quasi-static impurity distribution 
interacting with type-($1$) particles. 
Alternatively, one may concentrate on type-($2$) particles with charge $\sqrt{3}$,
which see type-($1$) particles as a distribution of charge $n/\sqrt{3}
\leq \sqrt{3}$ impurities. Therefore both types of particles are more strongly
repelled by those of their own species than by particles of different type. One
thus obtains a stable homogeneous mixture of two plasmas, which is shown in Fig.
\ref{n<3}(a).

\item {\sl Part II ($3<n<7$).} Although type-($1$) particles are still more
strongly repelled by those of their own species than by type-($2$) particles 
($n/\sqrt{7}\leq \sqrt{7}$), this is not the case for type-($2$) particles. Because
their inter-species repulsion is now weaker than that which they experience from
type-($1$) particles ($n/\sqrt{3}>\sqrt{3}$), they prefer to gather rather than
to be mixed to type-($1$) particles. This indicates a tendency to phase-separate, 
and the effect manifests itself in an unphysical negative filling factor 
$\nu_1$. The divergence of the filling factors at the artificial value of
$n_c=\sqrt{21}\simeq 4.6$ in Fig. \ref{73n} is due to the vanishing of the eigenvalue $\lambda_-$. Above $n_c$, the original minimum of the energy functional
(\ref{eq06}) evolves into a saddle point, and the first stability condition
of non-negative eigenvalues is no longer satisfied. Indeed, one notices that 
the filling factors interchange their role -- although the repulsion between 
type-($1$) particles is stronger than that between type-($2$) particles and one 
would therefore intuitively expect that $\nu_1<\nu_2$, one finds $\nu_1>\nu_2$
for $n>n_c$. 

\item {\sl Part III ($7\leq n$).} Above $n=7$, the inter-component
repulsion is stronger than that between particles of like type. The phase
separation between the two plasmas, which we have alluded to in the discussion
of part II, is well-pronounced. Due to this strong inter-component repulsion, 
the interface between the two plasmas needs to be minimized, and this results in
an inhomogeneous state of two spatially separated plasmas, as shown in 
Fig. \ref{n<3}(b). Should the partial densities not be fixed, i.e. particles could flip from $1$ to $2$ state, it is clear that the system would favor a distribution where only one type of particles would remain. The interface between the two plasmas disappears.

\end{itemize}

\begin{figure}[!ht]
\centering
\includegraphics[width=8cm,height=6cm]{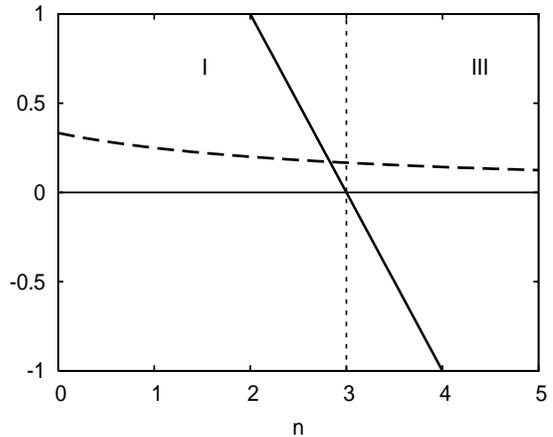}
\caption{\footnotesize Stability of the ($33n$) wavefunction. Equal filling factors, $\nu_1$ and $\nu_2$ (long dashed line), and the $\lambda_{-}$ eigenvalue (solid line) of the ($33n$) wavefunction are plotted as a function of $n$. For $n \leq 3$ (part I) the state is stable, as in Fig. \ref{73n}. There is no corresponding part II, since the densities never vanish and in part III the plasmas tend to phase separate.} 
\label{33n}
\end{figure}

Fig. \ref{33n} shows the stability graph for the a $(mmn)$ wavefunction
(here with $n=3$). In this case, the eigenvalues (\ref{eigen2}) and the
component filling factors (\ref{ff2}) become
\beq\label{eigenMMN}
\lambda_{\pm}=m\pm n \qquad {\rm and} \qquad \nu_1=\nu_2=\frac{1}{m+n},
\eeq
respectively. Although the component filling factors remain positive for all choices
of $n$, the eigenvalue $\lambda_-$ becomes negative for $n>m$, where one would
expect a phase separation between the two plasmas, as in the case of the
($73n$) wavefunction [Fig. \ref{n<3}(a)]. The critical value $n=m$ corresponds
to the Laughlin case with a SU(2) ferromagnetic spin wavefunction, discussed above.
For the case $n<m$, both
trial wavefunctions, ($mmn$) and ($nnm$), are valuable
candidates for the description of a potential FQHE at $\nu_T=2/(m+n)$ if only
the symmetry considerations for trial wavefunctions in the lowest LL are taken into
account. However, the plasma analogy indicates clearly that only one of the two
wavefunctions, namely ($mmn$), yields a stable physical state. At half-filling, 
e.g., the only SU(2) Halperin wavefunction which might yield a stable FQHE state
is ($331$), whereas $(113)$ corresponds to an unstable plasma, which is not 
evident from wavefunction calculations alone.\cite{Yoshioka}

		\subsection{The case $K=4$}
We now consider generalized Halperin wavefunctions with an internal SU($4$) symmetry. We restrict our studies to a particular subset of the latter, noted ($m_1m_2m_1m_2,n_en_+n_-$), the corresponding exponent matrices of which may be written as\cite{Goerbig_SU4}
\beq\label{M4}
 M_K =
\begin{pmatrix}
m_1&n_e&n_+&n_e\\
n_e&m_2&n_e&n_-\\
n_+&n_e&m_1&n_e\\
n_e&n_-&n_e&m_2\\
\end{pmatrix}.
\eeq
If applied to graphene, those correlation coefficients imply that one treats all intervalley components ($n_e$) on the same footing and intravalley ones separately ($n_+$,$n_-$ for $+$ and $-$ valley, respectively). This is an even more natural
assumption in the case of bilayer quantum Hall systems in semiconductor heterostructures, where interlayer correlations (described by the exponents
$n_e$) are weaker than intralayer ones ($n_+$ and $n_-$ which couple the different
spin orientations within the $+$ and $-$ layer, respectively).
Moreover, some intracomponent correlations are fixed to the same value for explicit calculation of the eigenvalues $\lambda$ and filling factors $\nu$ to be carried out. However, we will only settle here the conditions for all quantities to be positive (first and second stability arguments), and these conditions are satisfied if 
\beq\label{c4}
\left\{
\begin{array}{l}
m_1\geq n_+\\
m_2 \geq n_-\\
m_2+n_- \geq 2n_e\\
m_1+n_+ \geq 2n_e\ .\\
\end{array}
\right.
\eeq
The case where one of the two first inequalities, or the two last, turn into an equality corresponds to matrices of rank $r<4$. For $r=1$ Eq. (\ref{c4}) becomes an equation set and, once again, the only stable state is of Laughlin-type ($mmmm,mmm$) with SU($4$)-ferromagnetic ordering.
The stability criteria [Eq. (\ref{c4})] yield a slightly more complex interpretation: not only do intracomponent correlations have to be stronger than some intercomponent ones but there are also conditions between intercomponent coefficients. This may be more easily understood within the ($3535,n22$) state, where $n$ is left as a variable, similarly to the SU(2) case 
discussed in Sec. IV A. Again, we have chosen this state purely for illustration
reasons. 

\begin{figure}[!ht]
\includegraphics[width=8cm,height=6cm]{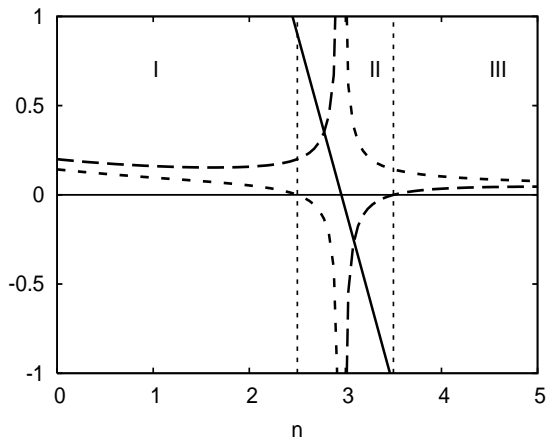}
\caption{\footnotesize Stability of the ($3535,n22$) wavefunctions. Filling factors $\nu_1$ and $\nu_3$ (long dashed line), $\nu_2$ and $\nu_4$ (short dashed line), and the third eigenvalue $\lambda_3$ (solid line) of the ($3535,n22$) generalized Halperin wavefunction are plotted as a function of $n$. For $n \leq 2$ (part I) all quantities are positive and the corresponding state is stable. Part II  corresponds to unphysical negative densities with a negative eigenvalue ($-0.08$). For $n \geq 4$, because of the eigenvalue $\lambda_3$ being still negative, the plasmas ($1$)-($3$) phase separate from ($2$)-($4$).}
\label{3535n22}
\end{figure}

Fig. \ref{3535n22} plots all filling factors and the relative signed eigenvalue $\lambda_3$. As for the case $K=2$, the graph is split into parts I,II and III.

\begin{itemize}
\item{\sl Part I ($n\leq 2$). } One notices that type-($1/3$) and ($2/4$) particles act identically therefore one can virtually treat the problem as for the case $K=2$. Type-($1/3$) and ($2/4$) plasmas are stable separately, with respect to the previous section on SU($2$) wavefunctions. Moreover, type-($1/3$) particles carry a $\sqrt{3}$ charge and are affected by type-($2/4$) quasi-static impurities of $n/\sqrt{3} \leq \sqrt{3}$ charge. Similarly, type-($2/4$) particles with charge $\sqrt{5}$ interact with type-($1/3$) plasma through a $n/\sqrt{5}\leq\sqrt{5}$ charge. Hence, one plasma suffers weaker repulsion from the other and type-($1/3$) and ($2/4$) will mix in order to form a stable homogeneous state.

\item{\sl Part II ($n=3$). }The $\lambda_3$ eigenvalue is now negative ($-0.08$) as well as some filling factors. In the plasma picture, type-($1/3$) particles are equally repelled by their own species and type-($2/4$) quasi-static impurities, each carrying a $\sqrt{3}$ charge. On the contrary, the type-($2/4$) plasma still experiences more "favourable" repulsion from type-($1/3$) particles. One cannot conclude about the stability at this level and some care has to be taken of the inner composition of the type-($1/3$) particles. Indeed, type-($1$) particles cary a $\sqrt{3}$ charge and interacts with type-($3$) quasi-static impurities via a $2/\sqrt{3}$ charge which is less than any other charge for this plasma. Similarly, type-($3$) will be less repelled by type-($1$) particles than by any other particles. Hence, type-($1/3$) plasma will tend to phase separate, which is contradictory to the phase mixing tendency of type-($2/4$) particles. As in the $K=2$ case, this yields unphysical negative densities for type-($1/3$) particles.

\item{\sl Part III ($4 \leq n$). } Above $n=4$, all filling factors are positive but $\lambda_3$ becomes more and more negative. For $n=4$, the same argument as above can be developped: type-($1/3$) plasma tend separate from type-($2/4$), whereas type-($2/4$) tend to mix with type-($1/3$). Surprisingly, this is not related to a negative density, as in any other case previoulsy discussed. This may be due to the composite nature of type-($2/4$) plasma. For $n\geq5$, both plasma are more severely repelled by each other such that the phase separation is complete.
\end{itemize}

		\subsection{Comparison with Exact-Diagonalization Studies}
The stability discussed so far is only related to the particular form of the wavefunctions and it has somehow to be  linked to the true ground-state of 
the quantum system with $N$ interacting electrons.
We therefore investigate the stability of generalized Halperin wavefunctions within exact-diagonalization studies. The system is mapped onto a sphere in the center of which a magnetic monopole is fixed to ensure a magnetic field orthogonal to the surface. This magnetic monopole creates $2S$ flux quanta threading the surface of the sphere. At a particular filling factor $\nu_T$, the relation between the number
of particles and that of flux quanta is 
$$2S=N/\nu_T-\delta,$$ 
where
\beq\label{FF}
\nu_T=\sum_{i,j}M_{ij}^{-1},
\eeq
and the shift 
\beq\label{shift}
\delta=\frac{1}{\nu_T}\sum_{i,j}M_{ij}^{-1}m_i
\eeq
is due to the finite-size geometry and depends on the particular wavefunction considered.

All calculations are performed within the lowest LL, with the help of Haldane's pseudopotentials,\cite{Haldane} $V_{ij}^l$, which determine
the interaction between two electrons of type $i$ and $j$ with a relative angular momentum $l$. Halperin's wavefunctions $(m_i,n_{ij})$ represent the exact ground-state for a model interaction such that
\begin{eqnarray}
V_{ii}^l&=&\left\{\begin{array}{l}1\;\;{\rm for}\;\;l < m_i\\0\;\;{\rm for}\;\;l\ge m_i\end{array}\right.\label{pseudopotentials}\\
V_{ij}^l&=& \left\{\begin{array}{l}1\;\;{\rm for}\;\;l < n_{ij}\\0\;\;{\rm for}\;\;l\ge n_{ij}\end{array}\right.\nonumber
\end{eqnarray}

One of the simplest non-trivial cases occurs when all intra-(inter-)component correlations are the same, i.e. $m_i=m$ and $n_{ij}=n$. These states are fully unpolarized and the corresponding filling factor and shift are
\begin{eqnarray}
\nu_T^{\{m,n\}}=\frac{K}{m+(K-1)n} &{\rm and}&\delta_{\{m,n\}} =m\, ,
\label{mnhalperin}
\end{eqnarray}
respectively. 
We have already shown in Sec. IV A and Sec. IV B that, within the plasma picture,
$m>n$ yields a stable and $m<n$ an unstable state, whereas $m=n$ represents 
a (stable) Laughlin state with SU($K$) ferromagnetic order. This stability 
criterion may also be obtained directly from the interaction model corresponding to
the $(m_i=m,n_{ij}=n)$-Halperin state: 
whenever this state is unstable, other zero-energy states with respect to the model interaction appear in the fully polarized sectors. These zero-energy states are 
quasihole excitations of the $(m_i=n_{ij}=m)$-Laughlin state the model interaction of which matches that of the Haperin state in that sector. This is a direct consequence of $\nu_T^{\{m,n\}}<1/m$ for unstable states, as may be seen from
Eq. (\ref{mnhalperin}) with $n>m$. Thus any Zeeman-type perturbation or any extra pseudopotentials beyond the model interaction dramatically change the polarization, as suggested by the plasma picture. For more generic Halperin wave functions, similar conclusions can be drawn when one of the $m_i$ is lower than $1/\nu_T$.

When the polarization is regarded as fixed, the phase separation is clearly observed through the study of pair correlation function, which are discussed in Sec. V B. As mentionned before, unstable systems will rather exhibit the instability through unphysical correlation functions.\cite{Forrester}

In addition to these general stability arguments, we investigate via exact diagonalization the $K=2$, ($331$) and ($113$) wavefunctions studied by Yoshioka {\sl et al.}\cite{Yoshioka}
The first state is realized for the particular model $V_{\ua \ua}^1= V_{\da \da}^1 = V_{\ua \da}^0=1$, all other potentials being zero, and  $2S=2N-3$. The 
second one is related to $V_{\ua \da}^0= V_{\ua \da}^1 = V_{\ua \da}^2=1$ and $2S = 2N - 1$.
For $N=6$ electrons, and correspondingly $9$ and $11$ flux quanta, exact-diagonalization calculations yield an energy gap of $0.8$ for ($331$) as compared to $0.01$ for ($113$). Hence, the ($113$) state 
has an energy gap which is almost two orders of magnitude smaller
than the characteristic energy, which is set to one. One may therefore expect that
the ($113$) state is much less stable than the ($331$) state, as indicated by the 
plasma analogy and the abovementioned argument. This is
indeed the case as may be seen when other pseudopotentials are chosen non-zero
in a perturbative manner. We choose, for this investigation, to vary continuously
$V_A^3\equiv V_{\ua\ua}^3=V_{\da\da}^3$ and $V_E^1\equiv V_{\ua\da}^1$ from zero to one, in the case of the ($331$) state, and 
$V_A^1\equiv V_{\ua \ua}^1= V_{\da \da}^1$ and $V_E^3\equiv V_{\ua \da}^3$ for ($113$).
Our exact-diagonalization results show
that the unpolarized state described by the abovementioned wavefunctions (there is an equal number of spin $\ua$ and $\da$) is conserved only for the ($331$) case. 
Moreover, 
there is indeed an instability of the ($113$) state such that even a small perturbation in the pseudpotentials ($V_{A}^3=0.1$ for example) completely polarizes the state, in agreement with the general polarization argument given
above.

\begin{figure}[!ht]
{\begin{flushleft}
(a)
\end{flushleft}}
{\centering
\includegraphics[width=8cm,height=6cm]{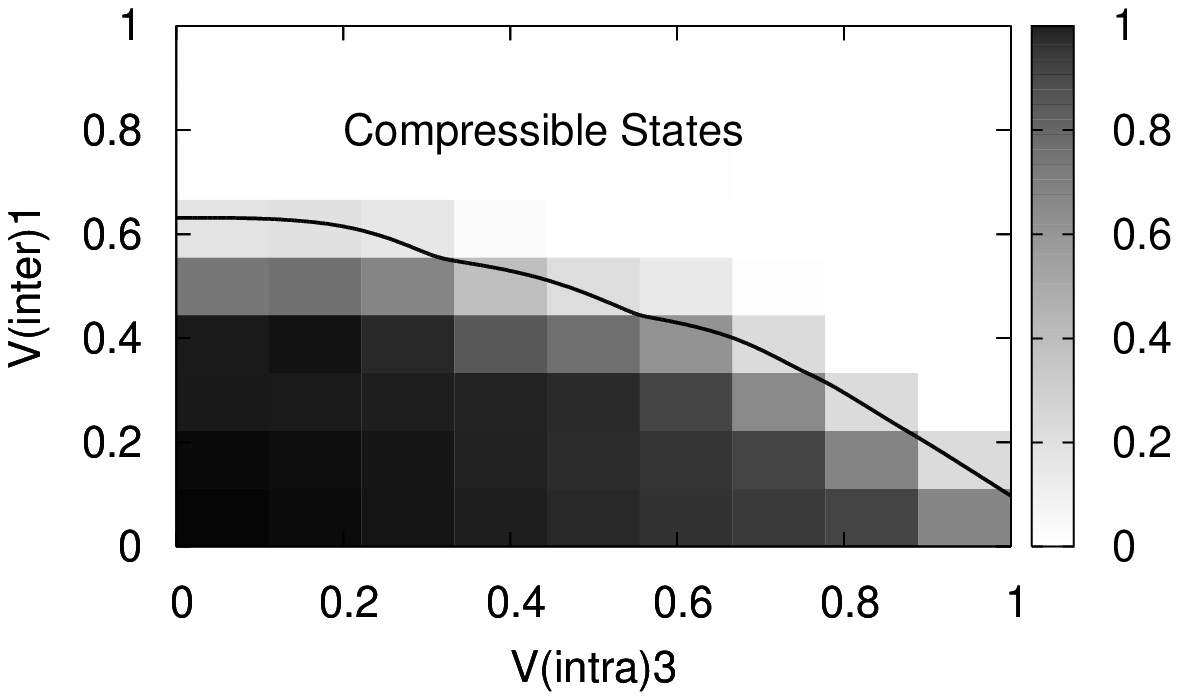}}

{\begin{flushleft}
(b)
\end{flushleft}}
{\centering
\includegraphics[width=8cm,height=6cm]{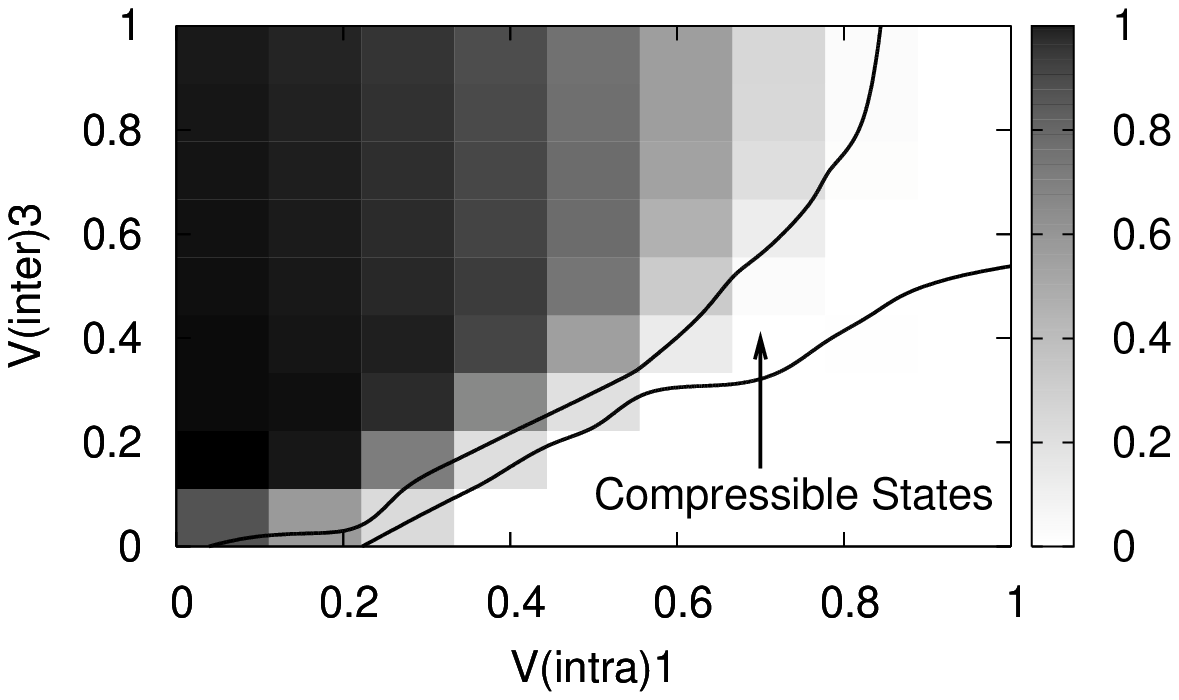}}
\caption{\footnotesize Phase Diagram for (a) 331 and (b) 113 ground-states -- Exact-diagonalization were performed with $N=6$ electrons and (a) $2S = 9$ (b) $2S = 11$ flux quanta. An unpolarized state is assumed. Pseudopotentials $V_{\ua \ua} = V_{\da \da}$ are noted $V(intra)$ whereas $V_{\ua \da}$ is $V(inter)$. The overlap between Halperin's wavefuncions (exact ground-state of the unperturbed model) and the exact-ground state is plotted in gray scale. We also indicate the separation between incompressible and compressible states (i.e. whether the ground state is in the $L=0$ sector or not) by a black line.}
\label{Phase2}
\end{figure}

We now focus on the unpolarized sector although the system is no more in its ground-state in the ($113$) case. The unpolarized sector may be physically relevant for a system with constrained polarizations such as for a bilayer configuration with equal densities in both layers. Fig.~\ref{Phase2} presents phase diagrams with compressible and incompressible states for ($331$) and ($113$) varying states. The overlap between the exact ground-state and the Halperin wavefunctions is represented by gradual shading. For the ($331$) case, it can be observed that there is a finite energy gap around the values $V_A^3=V_E^1=0$ of the exact model, and the overlap between the true ground state and the trial wavefunction remains large whenever the pseudopotentials $V_A^3$ and $V_E^1$ remain in this area. Further increase of the potentials leads to a gap collapse at relatively large values of
the pseudopotentials. Hence, this state is stable even in the case of more realistic interactions beyond the model situation.

In the case of the ($113$) wavefunction, the gap collapses rapidly even at small values of $V_{\ua \da}^3$ ($\approx 0.1$). A rather subtle perturbation would therefore completely change the system since a zero energy gap is incompatible with any FQHE. Outside the funnel-shaped area of compressible states, cf. Fig.~\ref{Phase2}(b), the overlap can be either quite small  ($\approx 0$, at larger values of $V_A^1$ and $V_E^3$) or quite large ($\approx 1$, in the vicinity of the model situation). This indicates that the incompressible states are described by states with different symmetry.
However, even though the system appears as stable up to relatively high $V_{\ua \da}^3$, the gap remains small ($0.01$ to $0.1$).
Moreover we emphasize that, in the case of an unfixed polarization, the ground state is no longer in the unpolarized sector.

Notice that the phase diagram in Fig. \ref{Phase2}(b) is not generic for all
unstable wavefunctions, but may 
be attributed to the pathologic model interaction of $(113)$.
General considerations on stability should only treat the polarization and pair correlation functions arguments.

In the same manner as for the case $K=2$, we compare the plasma picture with exact-diagonalization results for $K=4$.
For the special subset of matrices previously discussed, it has been checked numerically that unstable states are related to the existence of partially polarized zero-energy states with a lower number of flux quanta. As in the case $K=2$, those states will be favored when any Zeeman-like perturbation is introduced. The phase transition predicted within the plasma picture is recovered.
We checked this criterion for several particular wavefunctions, such as ($3333,233$) and ($3333,311$), and it appears that wavefunctions with unstable corresponding plasma do polarize, partially or completely, in agreement with the classical stability discussed in Sec. IV B.
For a given fixed polarization, phase separation is likely to be observed, as in the $K=2$ case.

	\section{Ground State Properties}

Although the plasma picture is a powerful tool for the study of intrinsic
properties of Laughlin and generalized Halperin wavefunctions, as shown in
the previous section, it gives in itself no indication of the physical state
chosen by the true interaction Hamiltonian. The Hamiltonian (\ref{eq05}) is 
obtained from a formal mapping of the wavefunctions to a corresponding plasma
model, but it is not related to the original Hamiltonian of interacting particles
in the lowest or partially filled higher LL. Indeed, incompressible quantum
liquids, which display the FQHE, are not found at all possible filling factors for
which one may write down a trial wavefunction. E.g. in the lowest LL, a Wigner
crystal is energetically favorable at $\nu<1/6.5$,\cite{lam} in the first excited
LL a succession of FQHE states and Wigner and bubble crystals\cite{goerbigRIQHE}
gives rise to a reentrant integral quantum Hall effect,\cite{eisenstein} and in 
even higher LLs stripe phases\cite{FKS,moessner} yield a highly anisotropic
longitudinal transport.\cite{expStripes} Which of these competing phases is 
indeed chosen depends on the precise form of the true interaction of electrons
in a fixed partially filled LL.

It is however possible to express the energy of Laughlin's wavefunction in
terms of the 3D Coulomb interaction potential $\mathcal{V}(\mathbf{r})= e^2/\epsilon r$
and the pair correlation function
$$g(r)\propto\int d^2z_3....d^2z_N\left|\Psi_m(z_1=0,z_2=r;z_3,...,z_N)\right|^2,$$
apart from a normalization constant,
\beq\label{egy}
E = \int d^2r \mathcal{V}(\mathbf{r}) \left[g(\mathbf{r})-1\right].
\eeq
The plasma picture may be of use here because the pair correlation functions for electrons and for plasmatic particles are the same, as may be seen from Eq. (\ref{eq02}). The pair correlation function may be
expanded as\cite{Girvin_Correlation}
\beq\label{gs}
\displaystyle{g (z)= 1-e^{-|z|^2/2}+\sideset{}{'}\sum_{n=1}^{\infty} \frac{2}{n!} \left( \frac{|z|^2}{4}\right)^n c_n e^{-|z|^2/4},}
\eeq
where the prime indicates a sum only over odd $n$ (due to Fermi statistics), and 
the expansion parameters $c_n$ vanish in the large-$n$ limit. These expansion
parameters are constrained in several respects. First, the short-range behavior $g \xrightarrow{|z|\rightarrow0}|z|^{2m}$ implies that $c_n = - 1$ for $n<m$.
Second, particular properties of the logarithmic potential in the plasma picture
can be used to derive sum rules, which act as further constraints.\cite{Baus,OCP,GMP,Kalinay}

		\subsection{Sum rules for SU($K$) pair correlation functions}

For wavefunctions with SU($K$) symmetry, there are $K(K+1)/2$ pair correlation functions if all densities are well-defined, i.e. if $M_K$ is invertible. The correlation function between type-($i$) and type-($j$) electrons is denoted by $g_{ij}(z)$, and one may generalize the expression (\ref{gs}) to the case
of $K$-component wavefunctions,
\beq
\displaystyle{g_{ij} (z)= 1-e^{-|z|^2/2}+\sideset{}{'}\sum_{n=1}^{\infty} \frac{2}{n!} \left( \frac{|z|^2}{4}\right)^n c_n^{(ij)} e^{-|z|^2/4},}
\eeq
where the expansion coefficients $c_n^{(ij)}$ vanish for large $n$. Here, the prime indicates summation over odd $n$ only for the intra-species functions $g_{ii}$. Indeeed, Fermi statistics is no more relevant when considering distinguishable electrons of type $i$ and $j \neq i$.
As for the Laughlin ($K=1$) case, the short-range behavior $g_{ij}(z) \xrightarrow{|z|\rightarrow0}|z|^{2n_{ij}}$ implies that $c_n^{(ij)} = -1$ for $n<n_{ij}$. 

Further sum rules may be derived within the picture of $K$ correlated
plasmas introduced in the previous section. In order to derive those in the
simplest manner, we decouple the different plasmas with the help of 
an orthogonal transformation on the densities,
\beq\label{basis}
 \begin{pmatrix}
   \rho_1' (\mathbf{r}) \\
   \vdots  \\
   \rho_K' (\mathbf{r})
 \end{pmatrix}
 =
 P
  \begin{pmatrix}
   \rho_1 (\mathbf{r}) \\
   \vdots  \\
   \rho_K (\mathbf{r})
 \end{pmatrix},
\eeq
which diagonalizes the exponent matrix, $M_K = P^{\top} \,D P$, 
in terms of the orthogonal matrix $P$.
The diagonalized Hamiltonian thus reads
\beqn
\nn
\Hmath[\{ \rho_i'(\mathbf{r}) \}] & = & \sum_{i=1}^K \Hmath^{(i)}[\rho_i'(\mathbf{r})]\ ,\\
\nn
\Hmath^{(i)}[\rho_i'(\mathbf{r})] &=&{ -\iint_{\Omega} d^2r \, d^2r'
   \rho_i' (\mathbf{r})
\frac{\lambda_i}{2}
\ln |\mathbf{r} - \mathbf{r'}|
   \rho_i' (\mathbf{r'})} \\
&& +\int_{\Omega} d^2r
   \rho_i' (\mathbf{r})
\frac{\alpha_i |r|^2}{4}\ ,
\eeqn
where $\lambda_i$ is the $i$-th eigenvalue of $M_K$ and $\alpha_i = \sum_j [P]_{ij}$. The Hamiltonian $\Hmath[\{ \rho_i'(\mathbf{r}) \}]$ is a sum of $K$ independent Hamiltonians $\Hmath^{(i)}[\{ \rho_i'(\mathbf{r}) \}]$, each of which corresponds to a single 2DOCP.
The correlation functions for these $K$ plasmas must therefore obey the usual sum rules for 2DOCPs\cite{Baus,OCP,Kalinay}
%
%
\begin{subequations}
\beqn
\label{M0}
\Mmath_0 &=& -1\\
\label{M1}
\Mmath_1 &=& -\displaystyle{\frac{4}{2 \pi \beta \lambda_i}} \\
\label{M2}
\Mmath_2 &=& -\displaystyle{\frac{64}{(2 \pi \beta \lambda_i)^2}(1-\frac{\beta \lambda_i}{4})}\\
\label{M3}
\Mmath_3 &=& -\displaystyle{6 \frac{(\beta \lambda_i -6)(8-3\beta \lambda_i)}{(\pi \beta \lambda_i)^3}}
\eeqn
\end{subequations}
for the different moments 
$$\mathcal{M}_m\equiv(\rho_i')^{m+1}\int d^2r\, r^{2m}[g_{ii}'(\br)-1].$$
Here, primes indicate quantities in the diagonal basis. Eq. (\ref{M0}) is due to
the charge neutrality of the system, Eq. (\ref{M1}) reflects its perfect-screening
property, and Eq. (\ref{M2}) is a compressibility sum rule. The third moment 
[Eq. (\ref{M3})] has no apparent physical interpretation.
Because the plasmas are decoupled in the diagonal basis, there are no correlations
between different plasmas, i.e. $g_{ij}'(\br)=1$ for $j \neq i$. 

The pair correlation functions $g_{ij}(\br)$ in the original basis may be obtained
from the $g_{ij}'(\br)$ with the help of the inverse orthogonal transformation.
It is useful to start from the definition of the structure factor in reciprocal space, which is related to the pair correlation function by Fourier transformation,
$$
S(\bk)-1=\rho \int d^2r e^{i\bq\cdot\br}\left[g(\br)-1\right],
$$
for the simplest $K=1$ case. It may also be expressed in terms of density
operators,
\beq
\rho \, \Omega \, \displaystyle{ S(\mathbf{k}) = \langle \rho (\mathbf{k}) \rho(\mathbf{-k}) \rangle - |\langle \rho (\mathbf{k})\rangle|^2 },
\eeq
where the quantities in brackets are averages with respect to the probability density function. In the case of $K\neq 1$, the structure factor has a matrix
form,
$$
\Omega\sqrt{\rho_i\rho_j}S_{ij}(\bk)=\langle \rho_i(\bk)\rho_j(-\bk)\rangle
-\left|\langle \rho_i(\bk)\rangle\right|\left|\langle \rho_j(\bk)\rangle\right|\ ,
$$
and the associated pair correlation functions $g_{ij}(\br)$ may be obtained
from those in the diagonal basis with the help of
$S_{ij} (\mathbf{r}) = \delta (\mathbf{r}) \delta_{ij} + (\rho_{i} \rho_{j})^{1/2} [g_{ij} (\mathbf{r}) - 1]$, 
\begin{widetext}
\beq\label{gg'}
\begin{pmatrix}
\rho_1^2 [g_{11}(\bold{r})-1] & \cdots & \rho_1 \rho_K [g_{1K}(\bold{r})-1] \\
\vdots & \ddots & \vdots \\
\rho_K \rho_1 [g_{K1}(\bold{r})-1] & \cdots &\rho_K^2[ g_{KK}(\bold{r})-1] \\
\end{pmatrix}
 =
\delta (\mathbf{r})
\left[
-
\begin{pmatrix}
\rho_1 & \cdots & 0 \\
\vdots & \ddots & \vdots \\
0 & \cdots & \rho_{K} \\
\end{pmatrix}
+P^{\top}
\begin{pmatrix}
\rho_1' & \cdots & 0 \\
\vdots & \ddots & \vdots \\
0 & \cdots & \rho_{K}' \\
\end{pmatrix}
P 
\right]
+
 P^{\top}
\begin{pmatrix}
\rho_1'^2 [g_{11}'(\bold{r})-1] & \cdots & 0 \\
\vdots & \ddots & \vdots \\
0 & \cdots & \rho_K'^2 [g_{KK}'(\bold{r})-1] \\
\end{pmatrix}
P \, .
\eeq
\end{widetext}
The sum rules (\ref{M0}) to (\ref{M3}) for the diagonal basis thus immediatly yield those for the 
$K$ correlated 2DOCP, due to Eq. (\ref{gg'}). For instance, the zeroth- and the  first-moment rules are
\begin{subequations}
\beqn\label{sum1}
\int d^2r \, \rho_i [ g_{ij}(\mathbf{r})-1] &=& -\delta_{ij}\\
\label{sum2}
\int d^2r \,r^2 \rho_{i} \rho_j[ g_{ij}(\bold{r})-1]&=&-\displaystyle{\frac{4}{2 \pi \beta} [M_K^{-1}]_{ij}} \ ,
\eeqn
\end{subequations}
respectively.
Eqs. (\ref{sum1}) and (\ref{sum2}) are generalizations of results, which have previously been obtained for $K=2$.\cite{Girvin_Correlation,Forrester}
Second- and third-moment rules may be derived in the same manner.

\subsection{Exact-diagonalization results for pair correlation functions}

\begin{figure}
\includegraphics[width=6cm,angle=270]{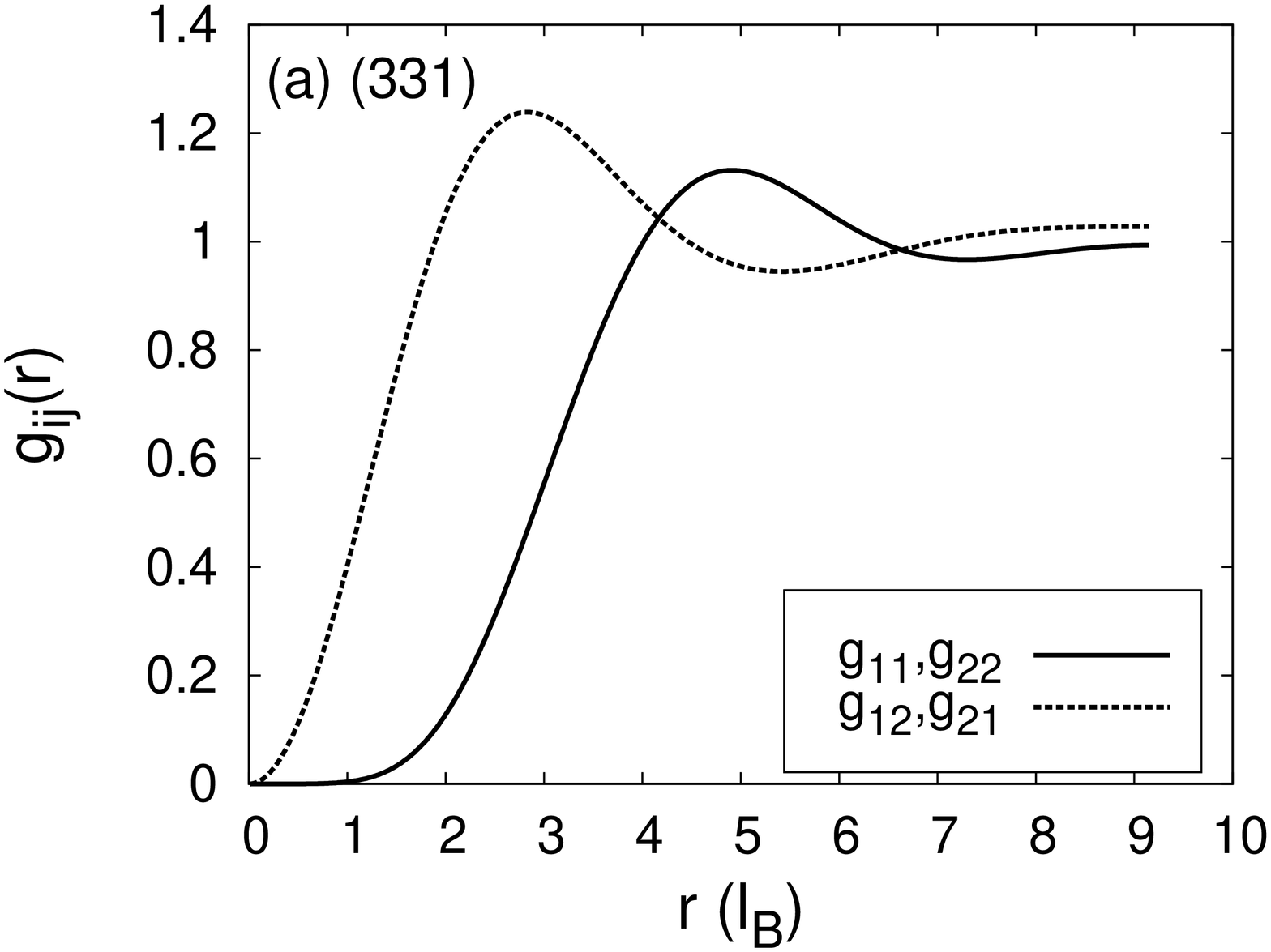}\\
\includegraphics[width=6cm,angle=270]{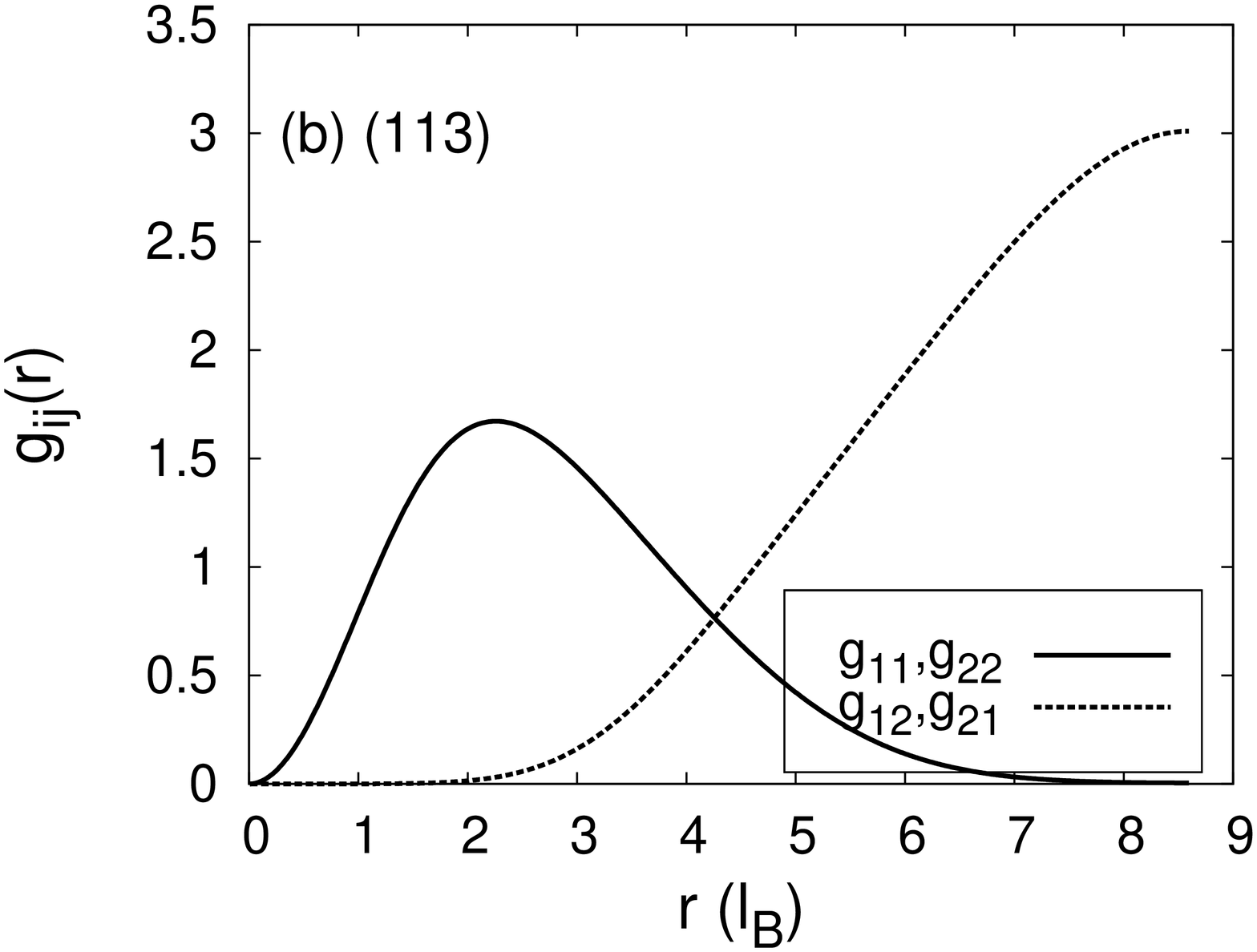}\\
\caption{\footnotesize Pair correlation functions related to (a) (331) for $N=10$ particles and (b) (113) states for $N=8$ particles, plotted as a function of distance (in units of magnetic length). Intra-component pairs ($g_{11}=g_{22}$) are plotted as a solid line and inter-component ($g_{12}=g_{21}$) as dashed ones. In the (331) graph, all functions go to 1 at infinity, which is typical of uniform density. On the opposite, the (113) intra-component function vanishes at large distances thus endowing a particles aggregate.} 
\label{corrSU2}
\end{figure}

The pair correlation functions of the ground-state may also be obtained from exact-diagonalization results. 
We first focus on the  SU(2) case by considering the same ($331$) and ($113$) wavefunctions 
discussed above and studied by Yoshioka {\sl et al.}\cite{Yoshioka} As mentioned
in Sec. IV C,
the first state is realized for the particular model $V_{\ua \ua}^1= V_{\da \da}^1 = V_{\ua \da}^0=1$, all other potentials being zero, and  $2S=2N-3$. The 
second one is related to $V_{\ua \da}^0= V_{\ua \da}^1 = V_{\ua \da}^2=1$ and $2S = 2N - 1$. Fig. \ref{corrSU2}(a) shows the ground-state pair correlation functions
for the $(331)$ state with $N=10$ electrons and $2S=17$ flux quanta, and the
results for the $(113)$ case ($N=8$ and $2S=15$) are displayed in Fig.
\ref{corrSU2}(b). 
The correlation function of a pair of electrons $(i, j)$ is the relative density of type-$j$ electrons when a type-$i$ electron is fixed at the origin, $\rho_{j}({\bf r})=\rho_{i} g_{ij}(\br)$. With this definition in mind, one can infer that the (331) state is rather well-behaved. Correlation functions vanish near the origin, as a result of repulsive interactions and Fermi statistics. The density peaks 
at finite distances indicate different layers of electrons (on average) with regular alternation of particles of type ($1$) and ($2$). The large-distance limit of type-$(1)$ particles is related to a uniform density. Hence, the system discussed is composed of a uniform and homogeneous mix of electrons of both types.
Unlike this first case, the $(113)$ state displays a vanishing intra-component function ($g_{11}=g_{22}$) at large distances. Electrons of both types thus tend to aggregate on a finite-size location. Moreover, the inter-component function ($g_{12}=g_{21}$) is maximum only at infinity which implies that the different types of electrons tend to spatially separate. This corroborates the plasma picture of phase-separated particles.

\begin{figure}
\includegraphics[width=6cm,angle=270]{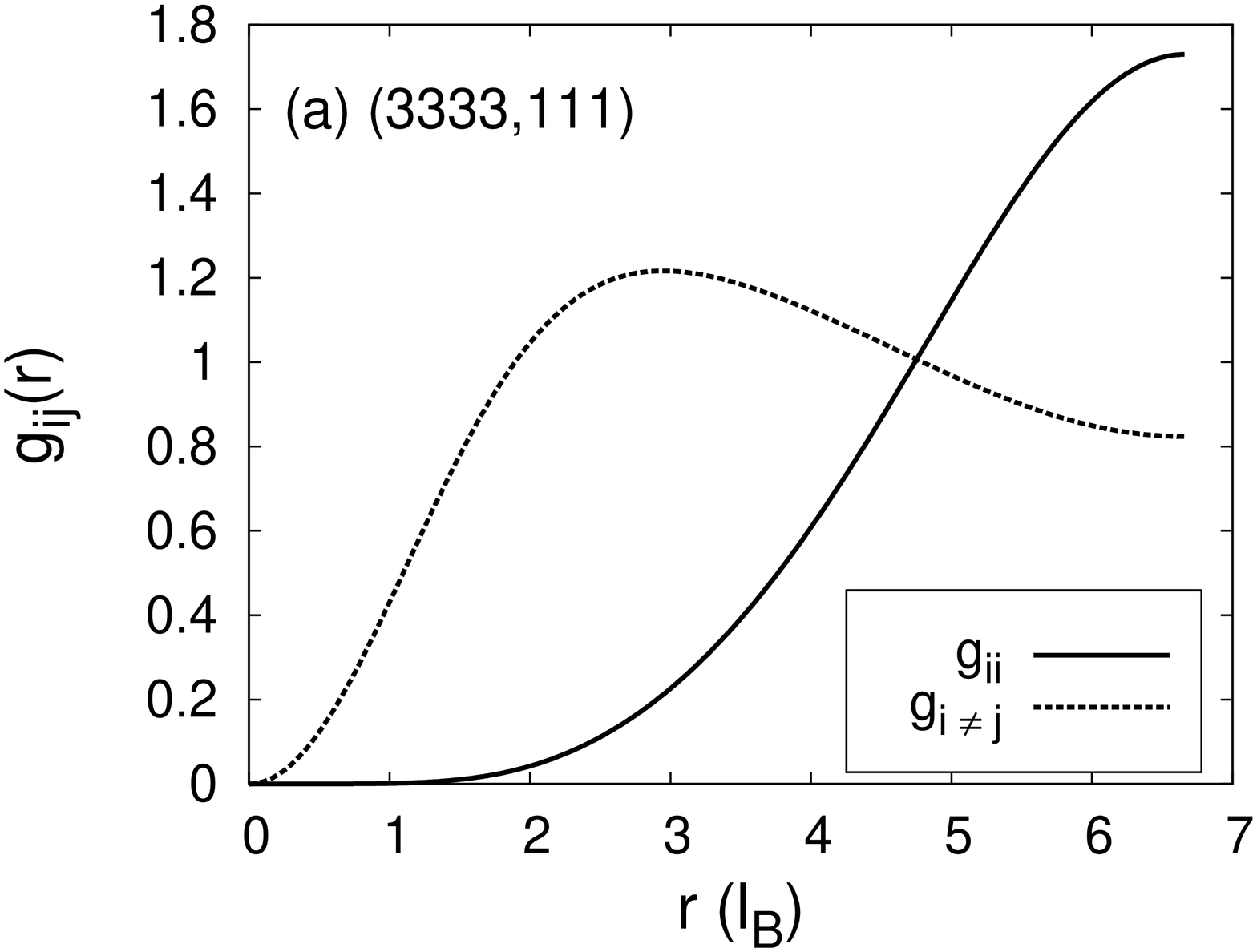}\\
\includegraphics[width=6cm,angle=270]{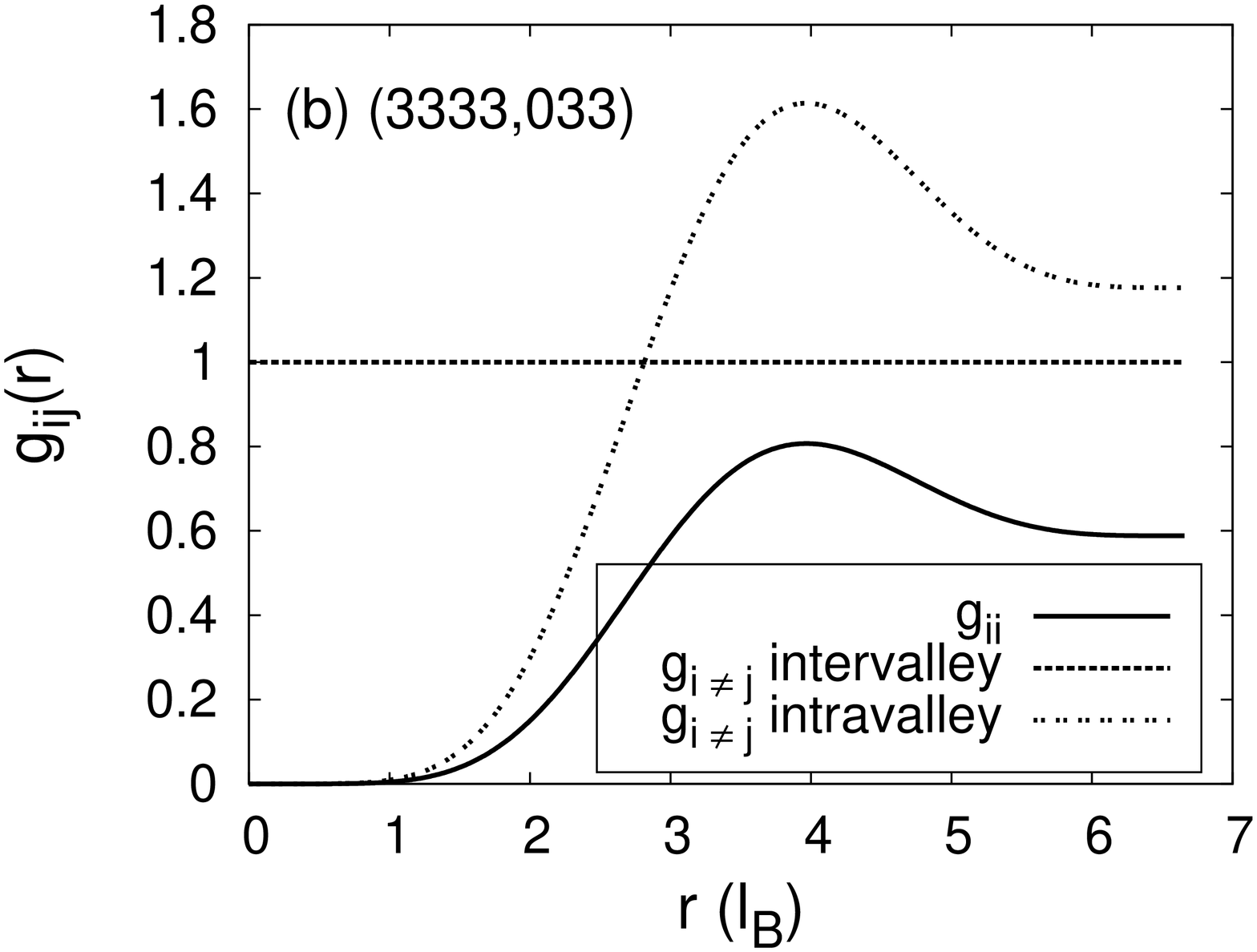}\\
\caption{\footnotesize Pair correlation functions related to (a) (3333,111) and (b) (3333,033) states for $N=8$ particles, plotted as a function of distance (in units of magnetic length). Intra-component pairs ($g_{ii}$) are plotted as a solid line. In the (3333,111) graph, the dashed line corresponds to the inter-component pairs ($g_{i \neq j}$, all being equal). In the (3333,033) graph, the dotted line shows the correlation function for electrons with opposite spin within any of each layer/valley. The dashed line corresponds to inter-component pairs between two different valleys or layers which are trivially constant in this non-correlated valley/layer example.} 
\label{corrSU4}
\end{figure}

We now turn to the SU(4) case by considering the two stable generalized Halperin wave functions studied in Ref. \onlinecite{Goerbig_SU4}, namely (3333,111) and (3333,033). The correlation functions are computed for $N=8$ electrons and $2S=9$ flux quanta. While the partial filling factors of the (3333,111) state are fixed and all equal to $1/6$, only the filling factor per layer is fixed for the (3333,033) ferromagnetic state. For a better comparison, we have set the partial filling factors to be the same as those of (3333,111). Thus in both cases, there are only two electrons per species leading to prominent finite size effect ($N=12$ is out of computational reach). The (3333,111) state is the straightforward SU(4) generalization of (331) and is therefore similar to its SU(2) counterpart (bearing in mind the small system size we are considering). The (3333,033) state consists of two independent spin insensitive $\nu=1/3$ Laughlin states in each layer. Both intra-layer and intra-component correlation functions are therefore identical to their Laughlin counterpart (up to normalization factors). The inter-layer pairs are trivially constant in this non-correlated layer (3333,033) state. 

Furthermore, we have confirmed numerically the validity of the sum rules
(\ref{sum1}) and (\ref{sum2}) for the SU(2) and SU(4) states discussed above. 
Notice that in the $(3333,033)$ state may be described alternatively by a 
SU(2) wavefunction, where the two components correspond to the two layer indices, 
regardless of the spin orientation. The relevant sum rules are therefore those of 
the SU(2) case, in which the exponent matrix $M_K$ in Eq. (\ref{sum2}) is 
invertible.

\section{Charged Excitations}

In this section, we study the charged excitations of SU($K$) Halperin wavefunctions 
with the help of the
plasma analogy. For $K=1$, quasi-hole excitations of the Laughlin wavefunction
may be written as
\beq
\displaystyle{\prod_{l=1}^N (z_l -z_0) \Psi_m(\{ z_k \})}.
\eeq
They consist of adding a zero of density, or a magnetic flux, in the electron liquid at the position $z_0$.

Within the plasma picture, the extra Jastrow term adds a new potential term to 
the Hamiltonian (\ref{eq03}),
\beq
\displaystyle{ \Hmath_N^{*}= \Hmath_N - \sum_{l=1}^N \ln|z_l-z_0|},
\eeq
which thus describes a 2DOCP with a fixed impurity at $z_0$ and charge $1/\sqrt{m}$. Because of the plasma's perfect screening ability,\cite{Bhatta} the particles are rearranged so that they screen the effect of the impurity in its vicinity. This requires $1/m$ particles in the plasma picture such that in the true electron liquid, the real charge $e^*$ of the excitation must be
\beq
\displaystyle{e^*=\frac{1}{m}},
\eeq
in units of the electron charge. Hence one electron can screen $m$ excitations.
			
We now investigate SU($K$) Halperin wavefunctions for which there exist different types of excitation. The quasi-hole wavefunction
\beq\label{excitew}
\displaystyle{ \prod_{k_i =1}^{N_i} (z_{k_i}^{(i)} - z_0) \times \Psi^{SU(K)}_{m_1,...,m_K;n_{ij}} }
\eeq
creates here an excitation of ($i$)-type electrons at the position $z_0$, i.e. adds a magnetic flux in the ($i$)-th component of the electron liquid.
The modified Hamiltonian in the plasma picture contains a new potential of an impurity that only affects particles of type ($i$),
$$
\Hmath_N^{(i)*}=\Hmath_N-\sum_{k_i=1}^{N_i}\ln \left|z_{k_i}^{(i)}-z_0\right|\ .
$$

Each of the $K$ correlated 2DOCP exhibits perfect screening ability, i.e. the plasma of type ($i$) must screen the impurity totally,
\beq\label{i1}
m_i e^{*(i)}_i + \sum_{j\neq i}n_{ij}e^{*(i)}_j=1,
\eeq
whereas plasmas of type ($j$), with $j\neq i$, screen a zero impurity
\beq\label{i2}
m_j e^{*(i)}_j + \sum_k n_{jk} e^{*(i)}_k = 0.
\eeq
Here, $e^{*(i)}_j$ is the quasiparticle charge, in units of the electron charge, carried by electrons of type ($j$) in the electron liquid for excitations in the ($i$) component. Eqs. (\ref{i1}) and (\ref{i2}) were previously derived for the $K=2$ case,\cite{DasSarma} and one may write them in a concise matrix form as
\beq\label{i3}
\sum_{k}n_{jk}e_{k}^{*(i)}=\delta_{ij} \qquad \Leftrightarrow\qquad
e_{j}^{*(i)}=(M_K^{-1})_{ji}.
\eeq
The last equation is valid only if $M_K$ is invertible. 
Indeed, if this is not the case, some component densities remain unfixed, as 
described in the previous section, and one could only consider excitation of groups of particles with a definite density. For instance, the ($mmm$) Halperin wavefunction has a non-invertible exponent matrix; the densities $\rho_1$ and
$\rho_2$ may fluctuate although their sum remains fixed, $\nu_T= 1/m$. Physical excitation must therefore not distinguish between the two components.

For the $K=4$ case, (3333,111) wavefunction\cite{Goerbig_SU4} exhibit four excitation types each of  which carries a $1/6$ charge, whereas for the (3333,033) wavefunction, only joined ($1$)-($3$) and ($2$)-($4$) excitations are to be considered, each of charge $1/3$.

Tab. \ref{tab01} shows examples of charged excitations for SU($2$) wavefunctions.
\begin{table}[!ht]
\begin{tabular}{|c|c|c||c||c|c|c|c|}
\hline
$m_1$ & $m_2$ & n & $\nu_T$ & $e_1^{*(1)}$& $e_2^{*(1)}$ & $e_1^{*(2)}$ & $e_2^{*(2)}$\\
\hline
3 & 3 & 0 & 2/3 & 1/3 & 0 & 0 &1/3 \\
\hline
3 & 3 & 1 & 1/2 & 3/8 & -1/8 & -1/8 & 3/8 \\
\hline
1 & 1 & 3 & 1/2 & -1/8 & 3/8 & 3/8 & -1/8 \\
\hline
3 & 3 & 2 & 2/5 &  3/5 & -2/5 & -2/5 & 3/5 \\
\hline
2 & 2 & 3 & 2/5 & -2/5 & 3/5 & 3/5 & -2/5 \\
\hline
3 & 3 & 3 & 1/3 & \multicolumn{4}{c|}{1/3}\\
\hline  
\end{tabular}
\caption{\footnotesize{Charged excitations of Halperin's wavefunctions}}
\label{tab01}
\end{table}
The first example describes two independent Laughlin states and it is consistent that a type-($1$) excitation should only affect type-($1$) particles.
The four next examples are related to the "($mmn$) and ($nnm$)" problem. One should notice that there are inconsistencies for ($113$) and ($223$) states. Indeed, when an extra flux quantum is added to the type-($1$) component, the number of flux quanta increases by one and the electron density remains the same. Therefore, the $\nu_1$ filling factor should decrease. However the total charge is conserved, so the sign of the quasi-hole charge should be the same as that of the electron, in order to compensate this "electronic" lack. This is not the case for ($113$) and ($223$) states which must be thus considered as unphysical, in addition to the
conclusions drawn in the previous sections.
The last example in Tab. \ref{tab01} is simply a Laughlin wavefunction split into two arbitrary sets. There is only one common excitation, as for the usual U($1$) case.

Similarly, Tab. \ref{tab02} shows some charged excited states for SU($4$) wavefunctions.
\begin{table}[!ht]
\begin{tabular}{|c||c||c|c|c|c|}
\hline
$m_1 m_2 m_1 m_2, n_e n_+ n_-$ & $\nu_T$ & $e^{*(1)}$& $e^{*(2)}$ & $e^{*(3)}$ & $e^{*(4)}$\\
\hline
3333,111 & 2/3 & 1/6 & 1/6 & 1/6 & 1/6\\
\hline
3555,222 & 2/5 & 1/5 & 1/15 & 1/15 & 1/15\\
\hline
3535,222 & 8/19 & 3/19 & 1/19 & 3/19 & 1/19\\
\hline
5555,222 & 4/11 & 1/11 & 1/11 & 1/11 & 1/11\\
\hline
\end{tabular}
\caption{\footnotesize{Charged excitations of SU($4$) Generalized Halperin's wavefunctions}}
\label{tab02}
\end{table}
All examples are associated with invertible matrices $M_K$. No proof is given here, but we conjecture that unstable states yield some inconsistencies concerning the charged excitations, as for the $K=2$ case.

	\section{Conclusions}

In conclusion, we have investigated the stability of Halperin wavefunctions
for $K$-component quantum Hall systems, with a particular emphasis on the cases
$K=2$ and $4$. The associated SU(2) and SU(4) internal symmetries happen to be 
the physically most relevant if one considers, e.g., bilayer quantum Hall
systems and graphene in a strong magnetic field. The $K=4$ case occurs when the
Zeeman effect is relatively small with respect to the leading interaction energy 
scales. In order to derive the stability criteria, we have generalized, in a 
systematic manner, Laughlin's plasma analogy to multicomponent systems. The 
validity of the criteria is corroborated with the help of exact-diagonalization
studies.

As for the conventional one-component quantum Hall system, the quantum-classical
analogy yields a compelling physical interpretation of the trial wavefunctions, in
terms of $K$ correlated 2DOCP. Besides the stability of the trial wavefunctions,
it also allows one to understand relevant ground-state properties, such as 
the associated pair-correlation functions, and fractionally charged quasiparticle
excitations. 

Whether the discussed trial wavefunctions correctly describe the true ground state
in physically relevant multicomponent systems, such as bilayer quantum Hall systems
or graphene, depends on the precise form of the interaction potential. The plasma
analogy with its rather artificial interaction may not give insight here, and variational or exact-diagonalization studies need to be performed to determine the correct ground state for a physical interaction potential. 
It has indeed been shown that
in the case of Coulomb interaction, a possible FQHE state at $\nu=2/3$ is not 
described by a generalized SU($4$) Halperin wavefunction.\cite{toke,Goerbig_SU4}
More complicated trial wavefunctions, such as composite-fermion\cite{toke}
or even more exotic states, 
may describe FQHE states at this and possibly other filling factors in a more
appropriate manner. However, in the U(1) one-component quantum Hall system, the 
inevitable starting point in the understanding of the FQHE is Laughlin's 
wavefunction;\cite{Laughlin} other wavefunctions may be viewed as sophisticated
generalizations of it. In the same manner, the study of SU($K$) Halperin wavefunctions and
the plasma analogy yield important physical insight into multicomponent
quantum Hall systems, and one may conjecture that they play a similar basic 
role for possible generalizations as Laughlin's in the U(1) case.

Furthermore, it has been shown in the SU(2) case, that not all possible, though
stable from our analysis, Halperin wavefunctions are valid candidates from a 
symmetry point of view. Indeed, most of the ($m,m,n$) wavefunctions are not
eigenstates of the total spin operator, the Casimir operator of SU(2), as they 
should for spin-independent interaction Hamiltonians.\cite{prange} The
(331) wavefunction discussed here, is, e.g., not an eigenstate of the total spin.
However, this problem may be cured by attaching the permanent of the matrix
$(z_i^{(1)}-z_j^{(2)})^{-1}$ to the (331) wavefunction.\cite{prange,HR} 
The situation is more complicated in the SU(4) case, where there are more
Casimir (spin-pseudospin) operators, and the wavefunctions should be eigenstates
of these operators. More detailed theoretical investigations are required to 
settle the question whether some SU(4) wavefunctions may be corrected in a 
similar manner.

	\section*{Acknowledgments}

We acknowledge fruitful discussions with J.-N. Fuchs, P. Lederer, R. Morf, and
S. H. Simon.
This work has partially been funded by the Agence Nationale de la Recherche
under Grant Nos. ANR-06-NANO-019-03 and ANR-07-JCJC-0003-01.

\end{document}